\documentclass[useAMS,usenatbib,usegraphicx]{mn2e}

\newcommand{\kev}{keV}
\newcommand{\fe}{Fe~K$\alpha$}
\newcommand{\etal}{et al.}
\newcommand{\threec}{3C~120}

\title[X-ray spectra of radio-loud AGN]
  {On the hard X-ray spectra of radio-loud active galaxies}
\author[D.\ R.\ Ballantyne, R.\ R.\ Ross \& A.\ C.\ Fabian]
  {D.~R.~Ballantyne$^1$\thanks{drb@ast.cam.ac.uk}, R.~R.~Ross$^2$ and
  A.~C.~Fabian$^1$\\
  $^1$Institute of Astronomy, Madingley Road, Cambridge CB3 0HA \\
  $^2$Physics Department, College of the Holy Cross, Worcester, MA 01610, USA}

\pagerange{\pageref{firstpage}--\pageref{lastpage}}
\pubyear{2002}

\usepackage{times}

\begin{document}

\label{firstpage}

\maketitle

\begin{abstract}
Over the last few years X-ray observations of broad-line radio
galaxies (BLRGs) by \textit{ASCA}, \textit{RXTE} and \textit{BeppoSAX}
have shown that these objects seem to exhibit weaker X-ray reflection
features (such as the iron K$\alpha$ line) than radio-quiet
Seyferts. This has lead to speculation that the optically-thick
accretion disc in radio-loud active galactic nuclei (AGN) may be
truncated to an optically-thin flow in the inner regions of the
source. Here, we propose that the weak reflection features are a
result of reprocessing in an ionized accretion disc. This would
alleviate the need for a change in accretion geometry in these
sources. Calculations of reflection spectra from an ionized disc for
situations expected in radio-loud AGN (high accretion rate,
moderate-to-high black hole mass) predict weak reprocessing
features. This idea was tested by fitting the \textit{ASCA} spectrum
of the bright BLRG \threec\ with the constant density ionized disc
models of Ross \& Fabian. A good fit was found with an ionization
parameter of $\xi \sim 4000$~erg~cm~s$^{-1}$ and the reflection
fraction fixed at unity. If observations of BLRGs by
\textit{XMM-Newton} show evidence for ionized reflection then this
would support the idea that a high accretion rate is likely required
to launch powerful radio jets.
\end{abstract}

\begin{keywords}
radiative transfer -- galaxies: active -- X-rays: galaxies -- X-rays:
general -- galaxies: individual: \threec
\end{keywords}

\section{Introduction}
\label{sect:intro}

In the standard disc-corona paradigm for the X-ray emission from
active galactic nuclei (AGN), the hard X-ray source is energized
within a magnetically dominated corona and illuminates the
optically-thick accretion disc below it (see \citealt{bel99} for a
recent review). The reprocessed emission from the disc imprints
spectral features on the observed spectrum, most notably the \fe\ line
(which may be broadened due to relativistic effects;
\citealt{fab89,tan95}; \citealt{fab00}) and a Compton reflection hump
at about 20--30~\kev\ due to the down scattering of higher energy
photons \citep[e.g.,][]{pou90}. By measuring the strength of these
reflection features and comparing to models of X-ray reflection, it is
possible, in principle, to constrain the solid-angle subtended by the
reflector (as seen from the hard X-ray source), and hence the geometry
of the accretion flow a few Schwarzschild radii away from the black
hole. Thus, measurement of the X-ray reflection signatures is not only a
valuable test for the disc-corona model, but also an important method
to probe the central engines of AGN.

Over the last few years observations of radio-loud AGN, particularly
broad-line radio galaxies (BLRGs), by \textit{ASCA} and
\textit{BeppoSAX} have indicated that these sources may have weaker
hard X-ray reprocessing features than radio-quiet AGN at similar
luminosities \citep*[e.g.,][]{eh98,gr99,esm00,gr01}. This has lead some
authors to speculate that the thin, optically-thick accretion disc in
radio-loud AGN changes to a hot, optically-thin flow before reaching
the black hole. Similar conclusions have also been made about Galactic
black hole candidates in their low/hard state: they also appear to
exhibit weaker reflection features than expected if the standard
accretion disc extends down to the last stable orbit
\citep[e.g.,][]{zds97,zds98,zds99,dz99}. Moreover, strong radio
emission from black hole candidates tends to be seen only when they
are in their low/hard state \citep{fen01}. A truncated disc in
radio-loud objects has also been supported by the recent theoretical
work of \citet{mei01} who considered the strength of the poloidal
magnetic field component expected in a standard \citet{ss73} disc as
compared to one in a ADAF-like flow \citep[e.g.,][]{ny95}. He
concluded that only the combination of a rapidly spinning black hole
and a geometrically thick accretion flow (i.e., an ADAF) could
produce the required jet power for a radio-loud AGN.

However, the problem of accurately determining the amount of
reflection in an X-ray spectrum depends on the sophistication of the
reflection models available to fit the data. Over the past 10 years
numerous calculations of X-ray reflection from Compton thick matter
have been performed to compare against the data
\citep*[e.g.,][]{ros93,zy94,mz95,ros99,nkk00,brf01}. These computations
have increased in sophistication over the years: first relaxing the
assumption of neutral material and then enforcing the condition of
hydrostatic balance on the illuminated material. 

Ionization effects can have an important impact on the shape of the
reflection spectrum and on the features imprinted on it
\citep{ros99}. In fact, if the disc surface is highly ionized the
reflection features become very weak, and if not taken into account,
\emph{a low reflection fraction can be measured}. Such ionized disc
models have been shown to account for the measured low reflection
fractions in Cyg~X-1 \citep{you01} and the black-hole transient Nova
Muscae \citep{dn01}. If this is the case, there is no need for a truncated
optically-thick accretion disc in these sources and any jet emission
would then arise from a standard disc that extends down to the last stable
orbit. However, the inner parts of the disc would be highly ionized,
likely due to extreme irradiation from the corona, and possibly
radiation-pressure dominated.

In this Letter, we investigate the idea that the low reflection
fractions measured in the hard X-ray spectra of radio-loud AGN are
simply the result of reprocessing from an ionized accretion disc. In
the next Section we calculate the reflection spectra from ionized
discs which may be appropriate for radio-loud AGN, and show that
they will exhibit weak reflection features. In Section~\ref{sect:res}
we apply ionized disc models to the \textit{ASCA} spectrum of the BLRG
\threec. Finally, we conclude by arguing that, in light of recent
results on the radio power of active galaxies, ionized accretion discs
should exist in most broad-line radio galaxies.

\section{Models of ionized reflection}
\label{sect:comp}

In this section we use the code presented by \citet{brf01} to compute
examples of reflection spectra from an irradiated disc within a
typical radio-loud AGN. Details of the numerical procedures may be
found in \citet{brf01} and references therein. The simulations model
the outer layers of an accretion disc which is illuminated by a
power-law continuum of X-rays. This external radiation extends from
1~eV to a sharp cutoff at 100~\kev\ and is parameterized by a
photon-index $\Gamma$ (so that photon flux$\propto E^{-\Gamma}$). The
gas structure is computed as a function of depth at a single point on
the disc. Once the irradiated material has relaxed into thermal,
ionization and hydrostatic balance, the reprocessed spectrum is
computed.

As discussed in Sect.~\ref{sect:discuss}, recent observational evidence
suggests that radio-loud AGN are likely accreting at a significant
fraction of their Eddington rate. Therefore, we have modelled a
radiation-pressure dominated disc accreting at 0.25 of its Eddington
rate. At these high rates, standard theory \citep{ss73}
predicts that the disc is puffed-up and the surface layers quite
tenuous, so the calculations covered the outer 15 Thomson depths of the
atmosphere in order to fully resolve the effects of the illumination.
The irradiating flux was chosen to be 10 times that of the soft flux
emitted by the disc (as might be expected from a magnetic flare event;
\citealt{nk01}), and strikes the surface of the atmosphere 4
Scharzschild radii from the black hole and at an angle of
$i=54.7$~degrees to the normal (so that $\cos i = 1/\sqrt{3}$, a crude
approximation to isotropic radiation). To simulate situations
appropriate for both the lower-luminosity broad-line radio galaxies
and radio-loud quasars, models were run with black hole masses of
5$\times$10$^7$~M$_{\odot}$ and 10$^9$~M$_{\odot}$.

\begin{figure*}
\begin{minipage}{180mm}
\centerline{
\includegraphics[angle=-90,width=0.5\textwidth]{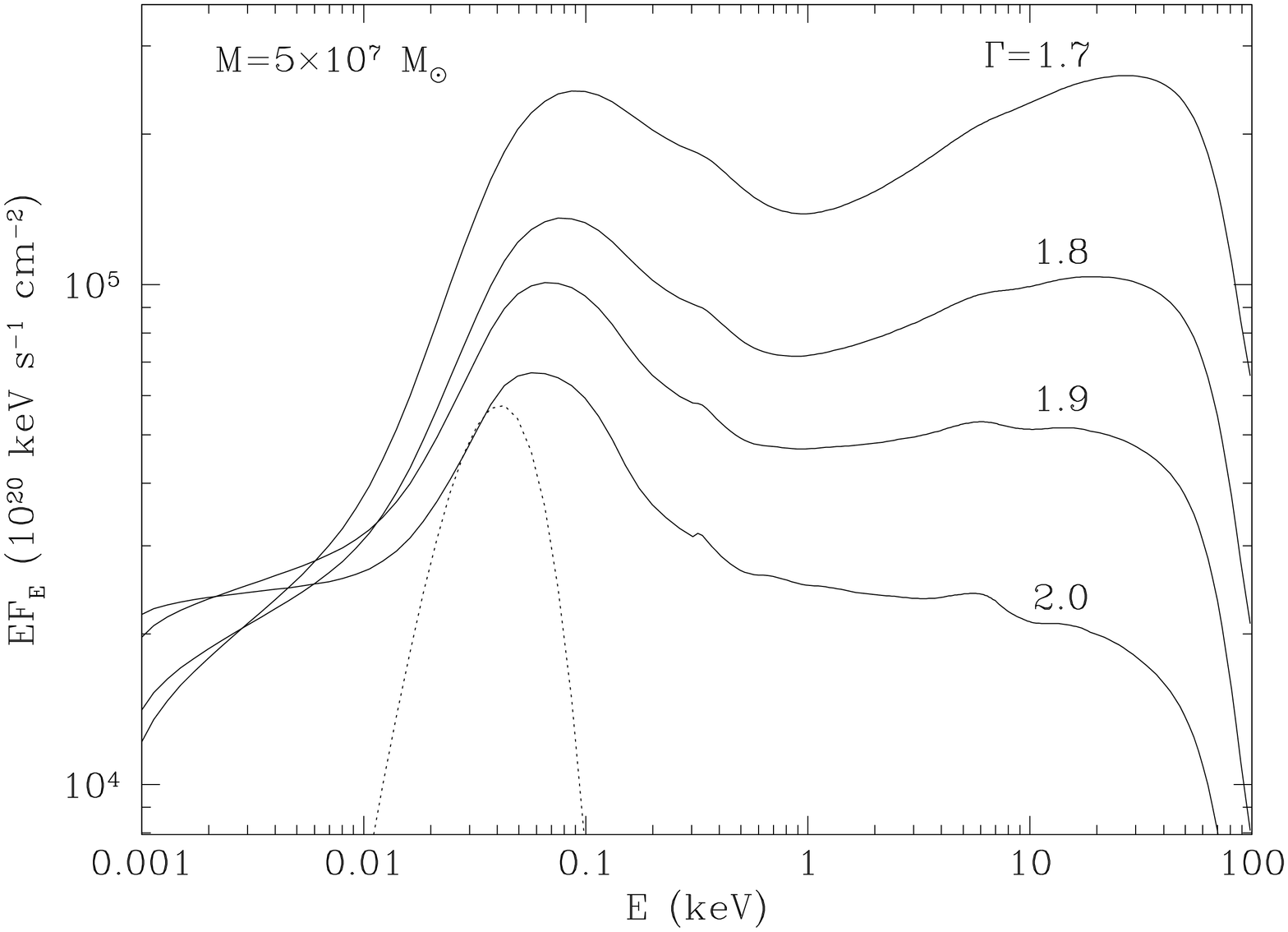}
\hspace{0.5cm}
\includegraphics[angle=-90,width=0.5\textwidth]{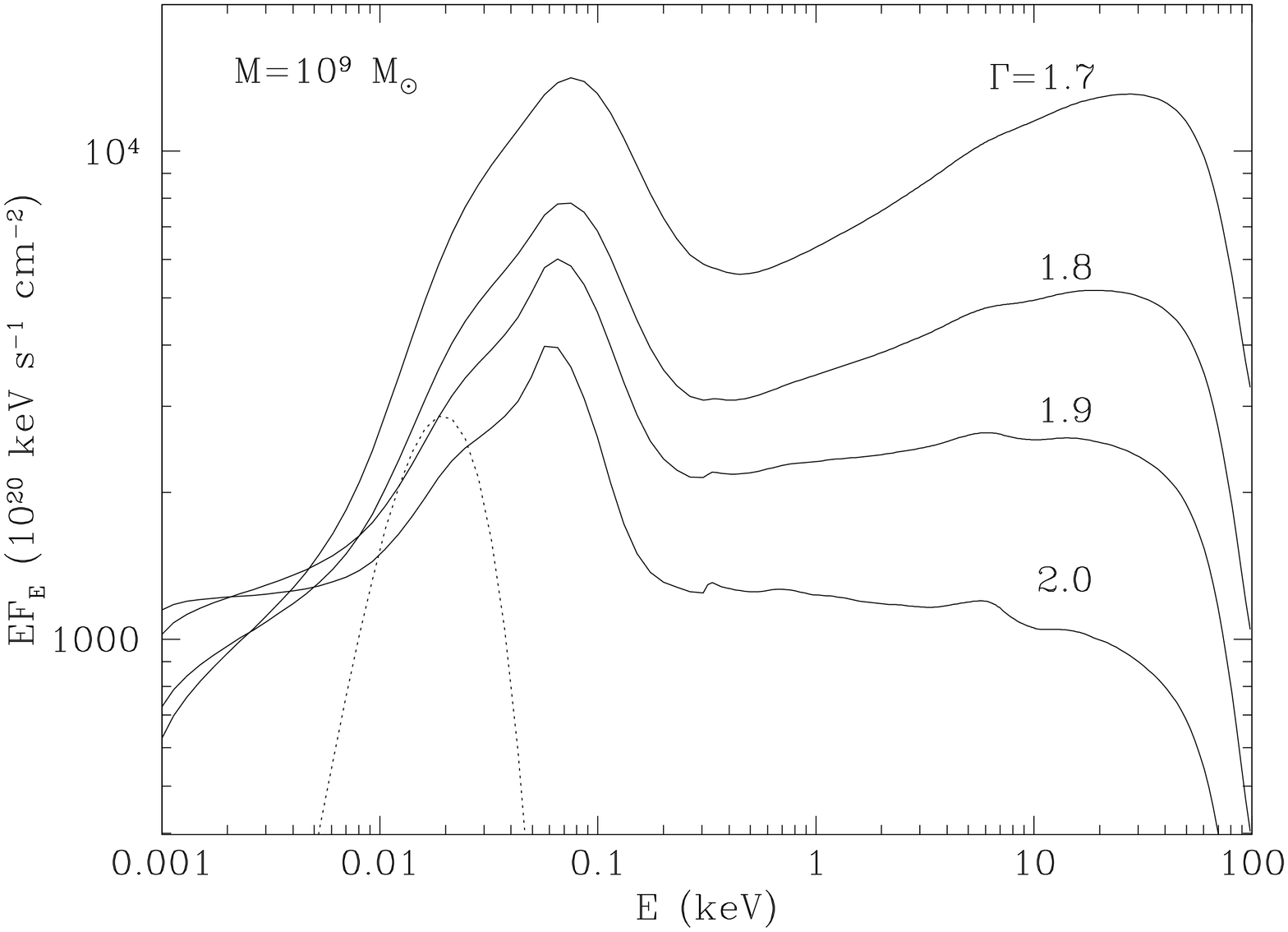}
}
\caption{Computed ionized reflection spectra from an irradiated
radiation-pressure dominated accretion disc accreting at 0.25 of its
Eddington rate. The disc is illuminated at a distance of four
Schwarzschild radii from the black hole with
$F_x/F_{\mathrm{disc}}=10$. (Left) Results for a
5$\times$10$^7$~M$_{\odot}$ black hole. (Right) Results for a
10$^9$~M$_{\odot}$ black hole. In both cases the spectra exhibit very
weak reflection features. The dotted lines denote the soft X-ray flux
from the disc incident on the bottom of the atmosphere. The spectra
have been vertically offset for clarity.}
\label{fig:spectra}
\end{minipage}
\end{figure*}

Reflection spectra were calculated for illuminating power-laws with
$\Gamma=1.7$, 1.8, 1.9 and 2.0, and they are shown in Figure~\ref{fig:spectra}.
The plots show that the reflection spectra exhibit very weak
reflection features as a result of the atmosphere being highly ionized
by the incident radiation. In fact, in all the calculations Fe is
fully stripped to at least 5 Thomson depths into the
atmosphere. Therefore, \fe\ photons have to traverse a
significant scattering layer before escaping. The Fe K edge feature also
suffers from Compton scattering and so the Compton reflection hump at
20--30~\kev\ becomes much less prominent. As was shown by \citet{dn01a} and
\citet{brf01}, when such highly ionized spectra are modelled by
neutral reflection models then low reflection fractions are easily
obtained. 

Although the X-ray data collected so far on broad-line radio galaxies
indicates that such extreme ionization is not needed (see
Sect.~\ref{sect:res}), it seems likely that weak reflection features
are to be naturally expected from a highly-accreting
object\footnote{Results for a wider range of accretion rates are
shown by \citet{brf01}, and are consistent with the ones presented
here.}. It should be pointed out, however, that this conclusion will
depend on the strength of the illuminating radiation and on the disc
structure. If the disc is weakly irradiated (i.e., the ratio of
incident flux to disc flux $\la$ 1) then there will likely be strong ionized
or neutral features in the reflection spectrum \citep{br02}. Also, at
high accretion rates the surface of the disc may become clumpy and
inhomogeneous which may enhance the strength of the reflection features
\citep{abm02}.

\section{Application of ionized disc models to 3C~120}
\label{sect:res}

\begin{figure*}
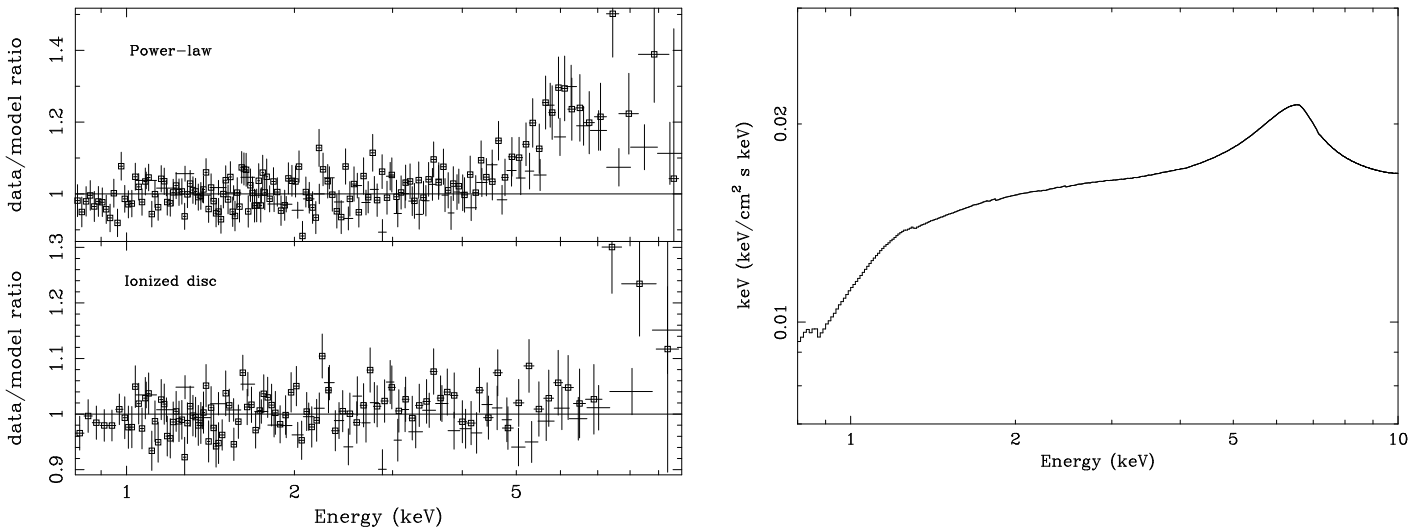

\begin{minipage}{180mm}
\centerline{
\includegraphics[angle=-90,width=0.5\textwidth]{ratio2.ps}
\hspace{0.5cm}
\includegraphics[angle=-90,width=0.5\textwidth]{asca_fit.ps}
}
\caption{(Left) Data-to-model ratios for both an absorbed power-law
fit (ignoring the data between 4 and 8~\kev) and the best fit ionized
reflection model to \threec. Open symbols denote data from the SIS-0
detector, while the points are data from the GIS-3. There does seem to
be some spectral hardening at high energies that is not explained by
the disc model. (Right) Ionized disc model of \threec\ as fit to the
1994 \textit{ASCA} data. The reflector is fairly highly ionized with
$\log \xi = 3.621^{+0.054}_{-0.049}$, so the \fe\ line arises from
He-like iron and has an EW of about 580~eV. The reflection fraction
was \textit{fixed} at unity, and solar abundances were assumed.}
\label{fig:3cmod}
\end{minipage}
\end{figure*}

To test whether ionized reflection models actually do fit the data of
a radio-loud AGN, we downloaded the \textit{ASCA} data of the BLRG
\threec\ from the \textsc{tartarus} database. \threec\ ($z=0.033$) is
the brightest BLRG in the X-ray sky ($F_{\mathrm{2-10~keV}} \approx
4.5 \times 10^{-11}$~erg~cm$^{-2}$~s$^{-1}$; \citealt*{sem99}), and
its X-ray properties are typical for its class, i.e.,
controversial. The 1994 50~ks observation of \threec\ by \textit{ASCA}
clearly showed an \fe\ line (see Fig.~\ref{fig:3cmod}), but despite
five different analyses of this dataset there has been no consensus on
the strength of the line. Workers who modelled the continuum as an
absorbed power-law \citep{rey97,sem99} consistently found the line to
be very broad ($\sigma > 1.5$~keV) and very strong (equivalent width
[EW] $\sim 1000$~eV). If, however, the continuum was modelled with
reflection \citep{gr97} or with a broken power-law \citep{woz98}, the
line width and EW dropped by over a factor of two. Recent observations
of \threec\ by \textit{RXTE} \citep{esm00} and \textit{BeppoSAX}
\citep{zg01} have provided some consistency by requiring reflection to
fit the continuum and obtaining an \fe\ EW $\sim 100$~eV. Thus,
previous work has shown that \threec\ has a curved X-ray continuum
with spectral hardening at high energies and an \fe\ line which may be
broad. 

The reflection models that have been previously fit to \threec\ have
assumed that the reprocessor is neutral and have all yielded
reflection fractions smaller than unity. This result, combined with
the weak \fe\ line, has been used as evidence for the claim that the
accretion disc is disrupted in radio-loud AGN \citep{esm00}. However, as
mentioned above, if the reprocessor is ionized then the reflection
features are naturally weak, and there is no need for a change in
accretion geometry. To test this hypothesis we fit the \textit{ASCA}
data of \threec\ with the constant density ionized disc models of
\citet{ros93} (see also \citealt{ros99}). Data between 0.8 and 10~keV
from both Solid state Imaging Spectrometers (SIS-0 \& SIS-1) and
between 1 and 10~keV from both Gas Imaging Spectrometers (GIS-2 \&
GIS-3) were fit simultaneously. The Galactic column of $1.23 \times
10^{21}$~cm$^{-2}$ was included in all fits.  The reflection spectrum
was relativistically blurred assuming that the emission arises between
10 and 1000 gravitational radii. The inclination angle was fixed at
14~degrees, as derived from observations of the radio jets
(\citealt{zen89}, but see \citealt{weh92} and \citealt*{walk88}). We
obtained a good fit ($\chi^2=1494$ for 1437 d.o.f.) with the following
parameters: $N^{\mathrm{intrinsic}}_{\mathrm{H}}=(5.92 \pm  0.89) \times
10^{20}$~cm$^{-2}$, $\log \xi = 3.621^{+0.054}_{-0.049}$, and
$\Gamma=1.875 \pm 0.015$, where $\xi$ is the ionization parameter of
the gas. The residuals to the fit and the ionized reflection model are
shown in Figure~\ref{fig:3cmod}.
The continuum parameters and EW of the \fe\ line ($\sim
580$~eV) are comparable to the results of \citet{gr97}. Most
importantly, our result is with the \textit{reflection fraction fixed
at 1}. This implies that there is no need for a truncated accretion
disc in \threec.

The \textit{RXTE} and \textit{BeppoSAX} observations of \threec\ both
suggest weaker \fe\ lines with EWs $\sim$ 100~eV
\citep{esm00,zg01}. While the \textit{ASCA} data analyzed here has
higher signal-to-noise and spectral resolution around 6~\kev\ than
these other data, it lacks the spectral coverage at energies greater
than 10~keV which is provided by these other telescopes. A comparison
between all three datasets is shown by \citet[][ their Figure 4]{zg01}. There
it can be seen that the shape of the \textit{ASCA} spectrum is consistent
with the \textit{RXTE} and \textit{BeppoSAX} data, suggesting that the
ionized disc model would be applicable to these other datasets.  The
difference in Fe line strength may be in part due to differences in
modeling the continuum. \citet{esm00} do not take in account ionized
reflection and \citet{zg01} cannot deal with highly ionized discs
(they assume a reflector temperature of 10$^5$~K, but the best-fit
slab has a surface temperature of 10$^7$~K). Another explanation
for the difference in EWs is that \threec\ was brighter by about 50
per cent (in the 2--10~\kev\ band) during the
\textit{RXTE} and \textit{BeppoSAX} observations than in the
\textit{ASCA} observation (again see Fig. 4 in \citealt{zg01}). 
If the \fe\ line flux remains relatively constant, as it is observed
to be in some Seyfert~1 galaxies (e.g., NGC~5548 \citealt{crb00};
MCG--6-30-15 \citealt{lfr00}), then the EW would be smaller in these
later observations.

The ratio plot shown in Fig.~\ref{fig:3cmod} indicates that there is
possible spectral hardening at high energies not accounted for by the
reflection model. This may be due to dilution from the radio jets
which could be particularly relevant for \threec\ as it has a
superluminal jet. Contamination from jet emission would also weaken
any reflection features. However, it is unclear whether or not the jet
dominates the high-energy emission of \threec. Although the object was
detected at $\sim 100$~\kev\ by \textit{OSSE} \citep{joh94}, it was
not picked up by either \textit{Comptel} or \textit{EGRET} at MeV
energies. At much lower energies, \citet{gr97} claimed to detect with
\textit{ROSAT} a variable soft-excess in \threec, as did \citet{zg01}
with \textit{BeppoSAX}. If \threec\ does have a soft-excess, then it
is unlikely jet contamination would be important at energies $<
10$~\kev, and an ionized disc is a valid model for the spectrum over
this band. A more precise determination of the spectrum of \threec\
requires an \textit{XMM-Newton} observation.

\section{Discussion}
\label{sect:discuss}

In this Letter we have presented the idea that the weak reprocessing
features seen in the \textit{ASCA} spectra of many radio-loud AGN are
the result of reflection off an ionized disc. With this explanation
there does not need to be a significant difference in the accretion
geometry between radio-loud and radio-quiet sources. This conclusion
is supported by evidence that the traditional bimodal distribution in
radio power of AGN is no longer valid. \citet{llr01} used results from
the FIRST Bright Quasar Survey to fill in the gap between the
radio-quiet and radio-loud populations, and concluded there was a
continuous distribution of radio-power that is correlated with both
black hole mass and accretion rate. Similarly, \citet{hp01} employed
high resolution radio and optical imaging of Seyfert galaxies to show
that many of these objects that were thought to be radio-quiet are
actually radio-loud (using the $R \equiv L_{\nu}\mathrm{(6\
cm)}/L_{\nu}\mathrm{(B)} > 10$ definition; \citealt{vis92}). These
authors also argued that there is a radio power-optical luminosity
correlation in active galaxies that stretches from Seyfert galaxies up
to luminous quasars.

There still remains the problem that only a small percentage of AGN
exhibit powerful, kpc-scale radio jets. The results of \citet{llr01}
suggest that a combination of high accretion rate and large ($>
10^{8.5}$~M$_{\odot}$) black hole mass is required to launch
ultra-relativistic jets. The radio luminosity-optical luminosity
correlations\footnote{The correlations seem to exist for both
radio-loud and radio-quiet objects, but with the slope being steeper
for the radio-loud ones.} discussed by \citet{hp01} also point toward
an accretion rate dependence. The actual physical mechanism
responsible for the radio emission and jets presumably is a
magnetohydrodynamic coupling between the accretion flow and a spinning
black hole \citep[e.g.,][]{bz77,mt82,ree82,bbr84,li00a,li00b}, but the
details are far from worked out
\citep[e.g.,][]{ga97,lop99,mei99,mei01}. However, the key parameters
must be the black hole mass and accretion rate because it is likely
that rapidly spinning black holes are found in many AGN, irrespective
of their radio power (e.g., MCG--6-30-15,
\citealt{iwa96,iwa99,wil01}). Indeed, studies of the hard X-ray
background suggest that most supermassive black holes are rapidly
spinning \citep{fi99,erz02}. Perhaps only rapid accretion onto a very
massive spinning black hole can provide the necessary energy to launch
kpc-scale radio jets.

However, observational evidence suggests that at least one more
parameter other than the black hole mass and the accretion rate is
needed to explain the triggering of powerful jets. Recent
\textit{XMM-Newton} observations of some high-luminosity radio-quiet
Seyferts have shown evidence for ionized \fe\ lines
\citep{ree01,pou01,orr01}, indicating that the accretion disc is
becoming more ionized as the luminosity and, likely, the accretion
rate increases \citep{br02}. A widening zone of extreme ionization on
the accretion disc would also explain the \textit{ASCA} observations
of radio-quiet quasars which found that the EW of the \fe\ line
becomes undetectable at a 2--10~\kev\ luminosity of
10$^{46}$~erg~s$^{-1}$ (the so-called X-ray Baldwin effect;
\citealt{it93,nan97,rt00}). Although most of these quasars have an
accretion rate close to Eddington and black hole masses $>
10^{8}$~M$_{\odot}$ \citep[e.g.,][]{llr01}, they do not possess
powerful radio jets. Therefore, additional quantities, such as the
magnetic field strength and configuration close to the black hole,
must also be important in determining the strength of the radio jets.
 
Nevertheless, it is possible that the weak X-ray reprocessing features
from radio-loud sources provide corroborating evidence that a high
accretion rate is important for the production of powerful
jets. Ultrasoft Seyferts (the subset of narrow-line Seyfert~1 galaxies
with unusual X-ray properties; \citealt{bra99,lei99a,lei99b,vau99b})
which are likely accreting close to their Eddington limit, have shown
evidence for ionized accretion discs in their X-ray spectra
\citep{com98,com01,tgn98,vau99a,bif01,tur01a,tur01b}. These objects
would not exhibit large scale radio jets because they do not have
massive enough black holes, or, perhaps, massive enough bulges to contain an
extensive reservoir of gas. Therefore, if the weak X-ray reflection
features in radio-loud AGN are due to ionization effects, then it is
likely to be a result of their high accretion rate, supporting the
view that a large accretion rate is required for the production of
relativistic jets. High quality spectral observations of radio-loud
AGN by \textit{XMM-Newton} will be able to test this hypothesis.

\section*{Acknowledgements}
This research has made use of the \textsc{tartarus} database, which is
supported by Jane Turner and Kirpal Nandra under NASA grants NAG5-7385
and NAG5-7067.  DRB acknowledges financial support from the
Commonwealth Scholarship and Fellowship Plan and the Natural Sciences
and Engineering Research Council of Canada. ACF and RRR acknowledge
support from the Royal Society and the College of the Holy Cross,
respectively.


\bsp 

\label{lastpage}

\end{document}